\documentclass[a4paper,fleqn]{article}
\usepackage[]{graphicx}

\newcommand{\bq}       {\begin{eqnarray}}
\newcommand{\eq}       {\end{eqnarray}}
\newcommand{\rf}[1]    {(\ref{#1})}

\newcommand{\en}       {\varepsilon}
\newcommand{\lp}       {\left}
\newcommand{\rp}       {\right}

\newcommand{\de}       {\partial}

\newcommand{\df}       {{\rm d}}

\newcommand{\ex}       {{\rm e}}
\newcommand{\im}       {{\rm i}}

\newcommand{\DR}       {{\rm DR}}

\newcommand{\ub}       {\bar{u}}
\newcommand{\vb}       {\bar{v}}
\newcommand{\wb}       {\bar{w}}

\newcommand{\tih}      {\tilde{h}}

\newcommand{\cL}       {{\cal L}}
\newcommand{\cM}       {{\cal M}}

\newcommand{\ol}[1]    {\overline{#1}}

\newcommand{\fh}       {{\textstyle{\frac 1 2}}}

\newcommand{\alt}      {\;{\tiny \parbox{3mm}{$<$\\$\sim$}}}

\begin{document}
\title{A perturbative model for predicting the high-Reynolds-number behaviour of the streamwise travelling waves technique in turbulent drag reduction}

\author{Marco Belan \and Maurizio Quadrio}

\date{}

\maketitle

\begin{abstract}

  The background of this work is the problem of reducing the aerodynamic 
turbulent friction drag, which is an important source of energy waste in 
innumerable technological fields (transportation being probably the most 
important). We develop a theoretical framework aimed at 
predicting the behaviour of existing drag reduction techniques when 
used at the large values of the Reynolds numbers $Re$ which are 
typical of applications. We focus on one recently proposed and very 
promising technique, which consists in creating at the wall 
streamwise-travelling waves of spanwise velocity (M.Quadrio, P.Ricco \& 
C.Viotti, {\em J. Fluid Mech.} {\bf 627}, 161-178, 2009).

A perturbation analysis of the Navier-Stokes equations that govern the 
fluid motion is carried out, for the simplest wall-bounded flow 
geometry, i.e. the plane channel flow. The streamwise base flow is 
perturbed by the spanwise time-varying base flow induced by the 
travelling waves. An asymptotic expansion is then carried out with 
respect to the velocity amplitude of the travelling wave. The analysis, 
although based on several assumptions, leads to predictions of drag 
reduction that agree well with the measurements available in literature 
and mostly computed through Direct Numerical Simulations (DNS) of the full 
Navier--Stokes equations. New DNS data are produced on purpose in this work to 
validate our method further.

The method is then applied to predict the drag-reducing performance of the streamwise-travelling waves  
at increasing $Re$, where comparison data are not available. The current belief, based on a $Re$-range 
of about one decade only above the transitional value, that drag reduction obtained at low $Re$ is deemed to 
decrease as $Re$ is increased is fully confirmed by our results. From a quantitative standpoint, however, 
our outlook based on several decades of increase in $Re$ is much less pessimistic than other existing estimates, and motivates further, more accurate studies on the present subject.

\end{abstract}

{\bf keywords}: Drag reduction, channel flow, moving walls, asymptotic expansions

\section{Introduction}

Driven by strong technological interest, the fundamental problem of manipulating turbulent flows is receiving more and more attention by the scientific community. Significant leaps forward in our physical understanding of turbulence and in our ability to simulate turbulent flows, either numerically or experimentally, have contributed in recent years to raise even further the interest in such topics, together with the growing concern with the energetic issue and environmental pollution. 

Perhaps the most difficult problem in this field is the control of turbulent wall flows to the aim of reducing the skin-friction drag. Applications (for example air-, water- and ground-based transport, as well as duct flows) are wide and the scientific challenge is significant, since wall flows, even in their simplest geometry, contain the essence of turbulence, i.e. the near-wall regeneration cycle \cite{jimenez-pinelli-1999,schoppa-hussain-2002} which adsorbs energy from the mean motion and redistributes it in an anisotropic way through the mediating action of pressure among turbulent fluctuations of the various velocity components.

A number of strategies to attack the problem exists, ranging from the so-called passive techniques, like riblets \cite{garcia-jimenez-2011-a} which modify the geometry of the planar wall in the hope of improving performance, to active, feedback-control techniques \cite{kim-bewley-2007} which build upon linear control theory to devise a control kernel capable of driving a large number of distributed microactuators on the basis of an input given by a large number of distributed microsensors. The simplicity of passive techniques is counterweighted by their so far limited performance; on the contrary, feedback control is performing well (in numerical simulations), but implementation issues are overwhelming. 

The best of both worlds could be found in the intermediate group of active, open-loop techniques. As for the passive techniques, they are reasonably simple to implement, not requiring distributed microsensors and microactuators. As for the feedback-control techniques, their global performance may be good enough to produce potential energy savings that are significantly larger that their working cost. One recent and promising strategy, especially so owing to its interesting performance in terms of net energy savings, has been described for ducts flows and boundary layers, and consists in creating at the wall a suitable distribution of spanwise velocity. The spanwise velocity varies sinusoidally in space as well as in time, to produce a spanwise-uniform wave that travels along the streamwise direction. This technique, introduced by Quadrio and coworkers in \cite{quadrio-ricco-viotti-2009} and reviewed by Quadrio in \cite{quadrio-2011}, can fully relaminarize the flow at low values of the Reynolds number $Re$, and yields net energy savings of more than 20\% for the higher $Re$ tested so far. An experimental verification exists \cite{auteri-etal-2010}, where drag reductions near to 50\% were measured in a low-$Re$ turbulent pipe flow.

As underlined in \cite{quadrio-2011}, one key question that has never been seriously addressed so far, however, is how these techniques behave when $Re$ is increased from the low values typical of the available experimental or numerical analyses to the high values typical of the applications. Direct Numerical Simulation (DNS) has been so far the numerical technique of choice for the minimum amount of flow modelling implied; however, it is obvious that, given the tremendous increase of its computing cost with $Re$, DNS will not allow us to get the much needed high-$Re$ information. Experiments here are stuck to an identical impasse, since the not-yet-satisfying development of suitable actuators limits the experiments to similarly low values of $Re$. Two scenarios are possible, and consistent with the limited data available, which evidence a mild decrease in performance (identified with the maximum obtainable drag reduction) over the extremely modest range of $Re$ explored so far (less than one decade above the subcritical value). One is, of course, that performances keeps decreasing as $Re$ increases, thus leading rapidly to a no-benefit situation well before the application-level $Re$ are reached. The other suggests that the observed decrease concerns the low-$Re$ regime only: at higher $Re$ performance stabilizes or at least decreases very slowly. This last scenario is consistent with the observations put forward in \cite{iwamoto-etal-2005} by Iwamoto et al., who studied the effect of completely removing the turbulent fluctuations in a thin near-wall layer, whose thickness was kept constant in wall units, and observed that the consequent reduction in skin-friction drag, after the initial drop at low $Re$, remains quite high even for high values of $Re$.

Early LES-based results are beginning to appear \cite{touber-leschziner-2012}, and seem to present analogous problems, providing reasonable results only when the employed spatial resolution becomes comparable to that of DNS. It becomes thus obvious that for understanding the high-$Re$ regime one has nothing left but resorting to the numerical solution of the Reynolds-averaged Navier--Stokes (RANS) equations closed with a suitable turbulence model. Unfortunately, no model exists to date that is capable of accounting properly for the new physics brought about by the wall-based control, inducing such large reductions of turbulent drag. 
The approach recently proposed by Moarref and Jovanovic \cite{moarref-jovanovic-2012} resorts to calculations based on RANS equations where an eddy viscosity is computed on the basis of spectral information of the non-controlled flow obtained by DNS. Though interesting in principle, and applied so far to the oscillating wall only, their approach presents the drawback of requiring prior DNS information, and thus precludes its applicability to the high-$Re$ regime of interest here.

In this paper, we follow an approach similar to \cite{moarref-jovanovic-2012} and 
present a predictor model that is based on the RANS equations and aims at predicting turbulent drag reduction without requiring high-$Re$ DNS information. Our model is specialized to the case of the streamwise-travelling waves, and how it performs in different situations still has to be verified. Moreover, it is a very simple model that makes several assumptions, some of which are known to be not entirely justified. It is however the best we can presently do by a perturbative approach, and we will demonstrate in the paper that, when properly developed by taking advantage of physical insight, the model provides us with useful results that may serve as an important guideline for improving our understanding of the problem at hand. The main assumption of the model is that the effect of the wall-based travelling waves is confined near the wall. This is a very reasonable assumption, as it is shown in \cite{quadrio-ricco-2011} that the waves produce a Generalized Stokes Layer (GSL), which is a generalization of the conventional Stokes layer produced by a wall in harmonic motion under a still fluid, and that the thickness of such GSL, compared to the distance between the channel walls, is extremely small when the waves produce drag reduction. Under this main hypothesis, we try and represent the effect of the GSL through a turbulent viscosity that depends on the wall conditions through a perturbative hypothesis that is applied both to the velocity field and to the turbulent viscosity itself.

The structure of this paper is as follows. \S\ref{PERTAPPROACH} describes the perturbative approach, with basic equations introduced in \S\ref{BASICEQNS}, the zeroth-order problem discussed in \S\ref{ORD0} and the first-order one in \S\ref{ORD1}; the main quantity related to relative drag reduction is discussed in \S\ref{DRAGREDUCTION}. Next, \S\ref{RESULTS} presents results of our predictive model and compares them with available DNS information. Then, \S\ref{USE} uses the predictive capabilities of the model to extract new information on the behavior of turbulent drag reduction, first in \S\ref{HIGHK} at high wavenumbers, and then in \S\ref{HIGHRE} at higher values of $Re$. Lastly, \S\ref{CONCLUSIONS} is devoted to a concluding summary.

\section{The perturbative approach}
\label{PERTAPPROACH}

\subsection{Basic equations}
\label{BASICEQNS}

The physical problem under study is characterized by a wall forcing, consisting in a spanwise velocity distribution which varies sinusoidally in space and time with a spatial scale $L_{wall}$ and a time scale $T_{wall}$. As a consequence, this distribution travels along the streamwise direction with a speed given by $L_{wall}/T_{wall}$. 
This symmetrical forcing does not introduce any net mean flow in the spanwise direction.

The definition of the mathematical model under consideration here relies first on the following definition for the mean value of a generic function of time
$f(t)$:
\bq
\bar{f} & = & \frac 1 T \int_0^T f \, \df t \;,
\label{MEAN}
\eq
where $T \ll T_{wall}$, i.e. $T$ must be large enough to smooth out the turbulent fluctuations, but it does not filter out the wall movement.
As we will see in the following, this hypothesis is justified as the travelling waves in their most interesting drag-reducing regime are characterized by a relatively long time scale compared to the turbulence time scale.

The basic equation system required here is the standard three-dimensional momentum and continuity equations set:
\bq
&&\de_t \vb_i + \vb_j \de_j \vb_i
= - \frac 1 \rho \de_i \bar{p} + \de_j \lp( \nu \de_j \vb_i - \ol{v'_i v'_j} \rp) \; 
\label{NSTURB}\\
&& \de_i \vb_i = 0
\label{DIVV}
\eq

\begin{figure}
\begin{center}
\includegraphics[width=0.7\textwidth,angle=0]{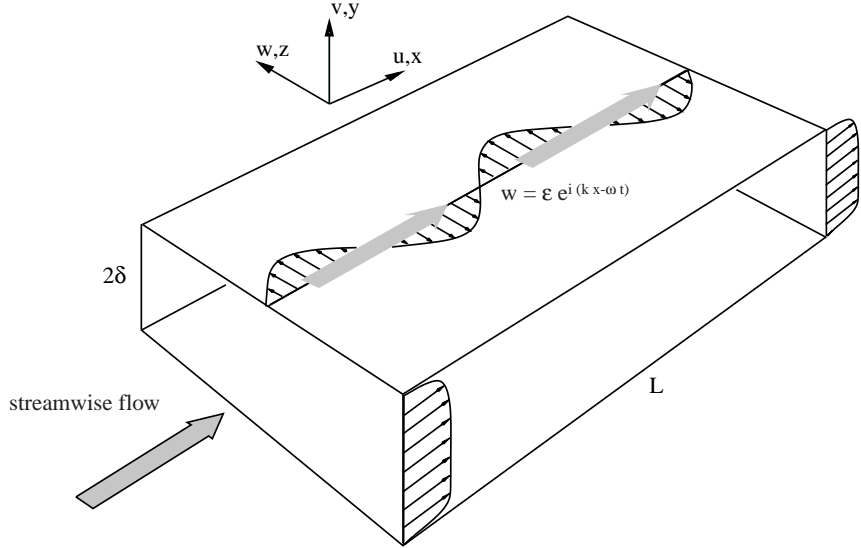}
\caption{Sketch of the coordinate system and the flow under study. The waves of spanwise velocity (with wavelength $2 \pi / k$ and oscillating period $2 \pi / \omega$) are applied at both walls and travel along the streamwise direction with speed $c = \omega / k$ that can be either positive (forward-travelling wave) or negative (backward-travelling wave). }
\label{flow} 
\end{center}
\end{figure}

The geometry and reference system are shown in figure \ref{flow}: $x$, $y$ and $z$ denote the streamwise, wall-normal and spanwise coordinate, with $u$, $v$ and $w$ the corresponding velocity components. We consider an indefinite turbulent plane channel flow, for which $x$ and $z$ are homogeneous directions, so that $\de_x$ and $\de_z$ are null for any velocity statistics. The flow is driven by a constant longitudinal pressure gradient $\de_x \bar{p} = G \rho < 0$, whilst the wall conditions are $\ub=\vb=0$ (and $\wb=0$ if no travelling waves are applied). 
The basic equations can be easily made nondimensional on the basis of the channel half-width $\delta$ and the bulk velocity. Here, to ease the comparison with \cite{quadrio-ricco-viotti-2009}, we choose as a reference velocity scale the centerline velocity $U$ of a Poiseuille laminar flow having the same mass flow as the channel under study. This velocity scale is related to the bulk velocity by a factor $3/2$.

After the introduction of the standard Boussinesq hypothesis, which expresses the Reynolds stresses as proportional to the gradient of the mean velocity field, the coefficient of proportionality being the turbulent viscosity:
\bq
 \ol{v'_i v'_j} = \frac 2 3 \kappa \delta_{ij}
- \nu_T \lp( \de_j \vb_i + \de_i \vb_j \rp) \; ,
\label{BOUS}
\eq
the equations \rf{NSTURB}, \rf{DIVV} become 
\bq
\de_t \ub + \ub \de_x \ub + \vb \de_y \ub
= - \frac 1 \rho \de_x \bar{p} 
+ \frac 2 3 \de_x \kappa +(\nu +\nu_T)(\de^2_x \ub +\de^2_y \ub) + 2(\de_y \nu_T) \de_x \ub + (\de_y \nu_T)(\de_y \ub +\de_x \vb) \; 
\label{UTURB}\\
\de_t \vb + \ub \de_x \vb + \vb \de_y \vb 
= \frac 2 3 \de_y \kappa+ (\nu +\nu_T)( \de^2_x \vb + \de^2_y \vb) + 2(\de_y \nu_T) \de_y \vb + (\de_x \nu_T)(\de_y \ub +\de_x \vb)  \; 
\\
\de_t \wb + \ub \de_x \wb + \vb \de_y \wb
= (\nu +\nu_T)( \de^2_x \wb + \de^2_y \wb) + (\de_x \nu_T) \de_x \wb + (\de_y \nu_T) \de_y \wb \; 
\label{WTURB}
\\
\de_x \ub + \de_y \vb  = 0 \; .
\label{CONT}
\eq

The perturbative hypothesis considered here expresses the mean turbulent flow as a basic part (the standard channel flow) plus a perturbation induced by the moving walls (a forcing acting on the channel flow) and can be written as: 
\bq
&&\ub = u_0(y) + \en\, u_1(x,y,t) + O(\en^2) \\
&&\vb = 0      + \en\, v_1(x,y,t) + O(\en^2) \\
&&\wb = 0      + \en\, w_1(x,y,t) + O(\en^2) \label{WPRT} \;,
\eq
where the perturbations are due to the boundary condition expressing the presence of the streamwise-traveling waves:
\bq
\wb(x,0,t) = \en \,\ex^{i (k x - \omega t)} = \en \ex^{i\theta} 
\label{WCOND}
\eq
that is $w_1(x,0,t) = \ex^{i\theta}$, scaled by $\en = w_{max}/U \ll 1$. Here the physical oscillation is represented by the real part of $\ex^{i\theta}$, and in what follows observable quantities are represented in general by the real parts of functions whilst the imaginary parts may contain related and/or redundant information. The small parameter $\en$  is assumed in this perturbative approach as the typical velocity amplitude of the superimposed $w$-perturbations, made dimensionless with respect to $U$. Given the typical values of $w_{max}$ described in \cite{quadrio-ricco-viotti-2009}, this assumption is not entirely justified, since $w_{max}/U$ can reach up to unity. As customary in asymptotic expansions \cite{vandyke-1975,kevorkian-cole-1981}, however, the small-amplitude assumption is useful to verify whether the analysis is able to yield reasonable results and predict important trends. 

The form of condition \rf{WCOND} suggests to hypothesize the functional form $\wb =\en\, f(y)\, \ex^{i \theta}+ O(\en^2)$ for $\wb$, and then to search for solutions of the kind 
\bq
&&\ub = u_0(y) + \en\, g(y)\, \phi_x( \theta ) \\
&&\vb = 0      + \en\, h(y)\, \phi_y( \theta ) \\
&&\wb = 0      + \en\, f(y)\, \ex^{i \theta} 
\label{WPERT}
\eq
at the first order in $\en$, where $f(y)$ is normalized to the $w$-perturbation small amplitude $\en$ at the wall, so that $f(0)=1$. 
The functions $\phi_i$ can be expressed as generalized Fourier series
\bq
\phi(\theta) = c_0 + c_1 \ex^{i \theta} + c_2 \ex^{2 i \theta} + ...
\eq
and the simplest model is obtained by truncation at the first order:
\bq
&&\ub = u_0(y) + \en\, g(y)\, (a_0 + a_1 \ex^{i \theta}) \\
&&\vb = 0      + \en\, h(y)\, (b_0 + b_1 \ex^{i \theta}) \\
&&\wb = 0      + \en\, f(y)\, \ex^{i \theta} .
\eq
By substitution in the continuity equation \rf{CONT} it can be easily shown that 
solutions of this kind can exist only if $b_0=0$. Coefficients $a_1,b_1$ can be factorized, so the
model takes the form
\bq
&&\ub = u_0(y) + \en\, g(y)\, (a + \ex^{i \theta}) \label{UPERT}\\
&&\vb = 0      + \en\, h(y)\, b\, \ex^{i \theta} \label{VPERT} \\
&&\wb = 0      + \en\, f(y)\, \ex^{i \theta} \label{WPERT2}
\eq
where the functions $f,g,h$ and the constants $a,b$ may depend in general on the parameters $k$ and $\omega$.
Since the forcing modifies the mean velocity gradients, the perturbative hypothesis can be extended 
in a similar way to the turbulent viscosity: 
\bq
\nu_T = \nu_0(y) + \en\, \nu_1(y)
\label{NPERT}
\eq
where also $\nu_1$ may depend on the parameters $k$ and $\omega$ defining the forcing.

Equating the terms of order $\en^0$ and solving the ensuing system leads to the determination of the basic flow, i.e. the channel flow without forcing, which will be briefly discussed in \S\ref{ORD0}. Solving the problem corresponding to higher order in $\en$ eventually provides us with the functions $f$, $g$ and $h$, so that the drag reduction effect can be discussed. This will be the subject of \S\ref{ORD1}.

\subsection{Order $\en^0$}
\label{ORD0}

Dimensional analysis on equations \rf{UTURB}--\rf{WTURB} may be carried out by considering at first the unperturbed channel flow velocity and length scales $U$, $\delta$ as in figure \ref{flow} together with the streamwise length $L$, such that $L/\delta \gg 1$. 
At the perturbative level, the motion has the new scales $\en$ for the velocity amplitude, and $T_{wall}\sim 1/\omega,\,L_{wall}\sim 1/k$ for the variations. We make the hypothesis that the wave velocity is of the same order of the stream velocity, i.e. $L_{wall}/T_{wall}\sim U$; again, this assumption is justified by considering the typical working conditions of the streamwise-traveling waves \cite{quadrio-ricco-viotti-2009}.
Then the growth order analysis leads to
\bq
&&u\hbox{-equation} = O(U^2/L) + O(\en) O(U^2 k)\\
&&v\hbox{-equation} = O(\delta/L)\,\lp[ O(U^2/L) +O(\en) O(U^2 k)\rp]\\
&&w\hbox{-equation} = O(\en)\, O(U^2 k)
\eq
Thus, at order $\en^0$ (unperturbed channel flow) the problem is properly described just by the dominant $O(U^2/L)$ terms of eq. \rf{UTURB}, whilst the higher order equations and terms will enter at the first order of approximation.

After substitution of equations \rf{UPERT}--\rf{NPERT} in the \rf{UTURB}--\rf{CONT},
the $\en^0$-terms of eq. \rf{UTURB} give 
\bq
-G + \nu_0'\,u_0' + \left( \nu  + \nu_0 \right) \, u_0'' =0
\label{U0}
\eq
(the symbol $'$ denotes $y$-derivatives) where $G$ is the longitudinal pressure gradient term, whilst at the same order the 
$v$-equation is negligible and the $w$-equation is trivially 0=0. 
Eq. \rf{U0} can be integrated giving
\bq
{A} - G \,y + \left( \nu  + \nu_0 \right) \, u_0' = 0
\label{UI}
\eq
where the constant $A$ is related to the wall condition, provided that $\nu_T(0)=\nu_0(0)=0$:
\bq
A = - \nu \, u_0'(0) , 
\label{C1}
\eq
i.e. $A$ is proportional to the basic wall stress. 
For the present purpose, it is sufficient to consider a single wall, setting there the origin of the $y$ axis.
The $\en^0$ problem is then closed by the boundary conditions $\ub=\vb=\wb=0$ at the wall and by the knowledge of one of the functions
$\nu_0(y)$, $u_0(y)$. The solution can be obtained by standard procedures, namely i) an hypothesis on the shape of $\nu_0(y)$ gives $u_0(y)$ by solving eq. \rf{UI} with $u_0(0)=0$; or ii) the knowledge of $u_0(y)$, from numerical simulations, experimental fits and/or special functions approximations, that leads to $\nu_0(y)$ through equations \rf{UI},\rf{C1}:
\bq
\nu_0 = \frac{G \,y + \nu  \lp[ \, {u_0}'(0) -  {u_0}' \rp] } { u_0' } .
\label{NUF}
\eq
The resulting channel flow (unperturbed, $\en=0$) will take the form
\bq
&&\ub = u_0(y) \\
&&\vb = 0      \\
&&\wb = 0      
\eq
where $u_0(y)$ represents the progressive velocity increase from the wall, loosing validity near the channel center where the derivative of the real velocity profile vanishes, a behaviour that cannot be reproduced here because eq. \rf{UI} degenerates for $u_0'(y)=0$. This is not a limit in this work, since the center region of the channel is not playing a key role in such wall-based drag reduction techniques that aim at modifying the near-wall turbulence regeneration cycle \cite{jeong-etal-1997}, and the drag reduction effects are supposed to depend on the interaction between the moving wall and these flow structures \cite{quadrio-2011}.

\subsection{Order $\en$}
\label{ORD1}

The dimensional analysis tells that the dominant momentum eqs at the $\en$ order are the $u-$ and $w-$ equations, that can be obtained looking
for $\en$-terms in the eqs \rf{UTURB},...\rf{CONT} after substitution of eqs \rf{UPERT},...\rf{NPERT}. The equation system to consider now consists of
the $u$-equation
\bq
\lp( 1 + a\,\ex^{-\im \theta} \rp)   \lp( \nu + \nu_0 \rp) \,g'' 
+ \lp( 1 + a\,\ex^{-\im \theta} \rp) \,\nu_0' \,g'
+ \lp( \im \,\omega - \im \,k\,u_0 -k^2\,\nu -k^2 \nu_0\rp)\, g +  \nonumber \\
+ \lp(\im \,k\, \nu_0' - u_0'\rp)\,b\, h 
+ \ex^{-\im \theta} ( \nu_1' u_0' +  \nu_1 \,u_0'') = 0 ,
\label{GEQ}
\eq
the $w$-equation
\bq
\lp( \nu  +  {\nu_0}  \rp) \,f'' +  {\nu_0}'  \, f' 
- \lp( k^2\,\nu   + k^2\, {\nu_0}  +  \im\,k\, {u_0} - \im \,\omega \rp)\, f = 0 
\label{FEQ}
\eq
and the continuity equation
\bq
g = \im \frac {b}{k} h' 
\label{CONT1}
\eq
(the effects of $k$ in the denominator will be discussed below).

Equation \rf{GEQ} involves the unknown functions $g,h$ and $\nu_1$. Substituting \rf{CONT1} in \rf{GEQ} the following equation for $h$ is obtained:
\bq
&&\ex^{-\im \theta}\,
\lp[\,
 \im \,ab\, \lp( \nu  + \nu_0 \rp) \,h'''
+ \im \,ab\,\nu_0'\,h'' 
+ k\,\nu_1'\,u_0' + k\,\nu_1\,u_0'' \,
\rp]+
\nonumber \\
&&+ \im  \lp( \nu  + \nu_0 \rp) \,h'''
+ \im \,\nu_0'\,h'' 
+ \lp( k\,u_0 - \omega -\im \,k^2\,\nu -\im \,k^2\,\nu_0  \rp) \,h' 
+(\im\, k^2 \nu_0'-k\, u_0')\,h = 0 \; .
\label{HEQ}
\eq
This equation is of the kind $\ex^{-\im \theta} [ \cL_1(y) h(y) + \cM_1(y) ] + [ \cL_2(y) h(y) ]=0$, 
where $\cL,\cM$ are linear operators, and proper solutions exist only if both parts vanish. The first part $\cL_1 h + \cM_1=0$ is
\bq
 \im \,a\,b\, \lp( \nu  + \nu_0 \rp) \,h'''
+ \im \,a\,b\,\nu_0'\,h'' 
+ k\,\nu_1'\,u_0' + k\,\nu_1\,u_0'' \, = 0,
\eq
and can be integrated to give 
\bq
B + k\,\nu_1\,u_0' + \im \,ab\,(\nu + \nu_0) \,h''=0
\eq
($B$ is a constant). This relation is satisfied by 
\bq
\nu_1 = -\frac{B + \im \,ab\,(\nu +\nu_0) \,h''}{k\,u_0'} ,
\label{NU1}
\eq
and the value of $B$ can be obtained through the condition $\nu_1(0)=0$.

The second part $\cL_2 h=0$ of eq. \rf{HEQ}, namely 
\bq
\im  \lp( \nu  + \nu_0 \rp) \,h'''
+ \im \,\nu_0'\,h'' 
+ \lp( k\,u_0 - \omega -\im \,k^2\,\nu -\im \,k^2\,\nu_0  \rp) \,h' 
+(\im\, k^2 \nu_0'-k\, u_0')\,h = 0,
\label{HEQ2}
\eq
can be solved for $h$ with boundary conditions $h(0)=h'(0)=0$, i.e. $u_1=v_1=0$ at the wall, together with
a third condition. This gives
\bq
h = h\,(y;\, k,\, \omega,\,s)
\eq
i.e. the function $h$ depends on $y$, on the parameters $k$ and $\omega$, and on a integration constant $s$.

The $w$-equation \rf{FEQ} may be solved for $f$ with condition $f(0)=1$, i.e. $\wb=\en\,w_1=\en\, \ex^{\im \theta}$ at the wall after eq. \rf{WPRT} and \rf{WPERT}, together with a second condition.

We note here that, for a laminar flow, the equations for $u$ and $w$ become simpler. In particular the laminar $u$-equation \rf{HEQ} reads
\bq
\ex^{-\im \theta}\,
 \im \,a\,b\, \nu \,h'''
+
\im \, \nu  \,h'''
+ \lp[ (k\,u_0 - \omega) -\im \,k^2\,\nu \rp] \,h' 
-k\, u_0'\,h = 0 \; 
\eq
i.e. takes the form $\ex^{-\im \theta} [\cL_1(y) h(y)] + [ \cL_2(y) h(y) ]=0$. Hence, the first part $\cL_1(y) h(y)=0$ must be satisfied by a trivial condition, i.e. $a=0$ and/or $b=0$; this condition will be identified in the next section. The second part $\cL_2(y) h(y)=0$ could be solved for $h$ as in the turbulent case, with the same boundary conditions, but in the next section it will be shown that the solution of this equation is not necessary in the laminar case.

For a laminar flow, even the $w$-equation \rf{FEQ} becomes simpler: 
\bq
\nu \,f'' - \lp[ k^2\,\nu   +  \im ( k\, {u_0} - \omega) \rp]\, f = 0 
\label{FLAM}
\eq
and admits decaying solutions in terms of confluent hypergeometric functions. For a very short $y$-domain, where the parabolic expression of $u_0$ (the Poiseuille velocity profile) can be linearized, the solutions take the form of Airy functions, in agreement with the analytical solution for the laminar GSL determined in \cite{quadrio-ricco-2011}.

\subsection{Drag reduction}
\label{DRAGREDUCTION}

The total wall stress is, after equation \rf{UPERT},
\bq
\tau_0 = \rho \lp[ \nu + \nu_T(0) \rp]\, (\de_y\ub)_{y=0} 
= \rho\, \nu \lp[ u'_0(0) + \en\, g'(0) \, (a + \ex^{i \theta}) \rp] ,
\eq
and in nondimensional (macroscopic) form:
\bq
c_f = \frac{\tau_0}{\fh \rho U^2} 
= \frac {2 \nu}{U^2} \lp[ u'_0(0) + \en\, g'(0) \, (a + \ex^{i \theta}) \rp] .
\eq
The drag-reducing effect of the travelling waves can be estimated by calculating the
mean value $c_{f,m}$ of the friction coefficient $c_f$ over a large time $T_g$ such that $T_g \gg T_{wall}$, this quantity is 
\bq
c_{f,m} = \frac {2 \nu}{U^2} \lp[ u'_0(0) + \en\, a\, g'(0) \rp]\,.
\eq
The unperturbed wall stress is
\bq
c_{f0} = \frac {2 \nu}{U^2}\,  u'_0(0) ,
\eq
so that the relative drag reduction, after equation \rf{CONT1}, is:
\bq
\DR = 1- \frac{c_{f,m}}{c_{f0}} 
= - \frac{\en\, a\, g'(0)}{u'_0(0)}
= - \frac{\im\, \en \, a\, b \, h''(0)}{k\, u'_0(0)} ,
\label{DR}
\eq
which means
\bq
\DR(k,\,\omega) \propto \, \frac {C}{k} \, h''(0;\,k,\,\omega,s)
\label{DRAGV}
\eq
where $C=ab$. The relation \rf{DRAGV} allows a map of the relative drag reduction to be drawn as a function of $k,\omega$ and of an integration constant $s$, which has the meaning of an accessory condition (obviously, the dependence of drag reduction upon the amplitude of the spanwise forcing cannot be determined by this perturbative approach). The factor $C$ cannot be determined at this level of approximation, but the required regularity as $k\to 0$ of $g'(y)$ and $\DR(k,\,\omega)$, 
both proportional to $b/k$ after eq. \rf{CONT1} and \rf{DRAGV}, suggests to assume 
\bq
b(k)\sim k^n \hbox{ with } n \ge 1 \hbox{ as } k\to 0 ,
\label{KT0}
\eq
an hypothesis which can be formally introduced in eq. \rf{UPERT}, {\em a priori}.  

It is important to note here that the physical meaning of $\DR$ is expected to be found in its real part, which depends both on the real and imaginary parts of $h$ since the factor $C$ is in general a complex quantity. The real part of eq. \rf{DRAGV},
writing $C=C_r+\im\, C_i$ and $h=h_r+\im \,h_i$, becomes
\bq
\DR_r(k,\,\omega) \propto \, \frac 1 k \lp[\, C_r \, h_r''(0;\,k,\,\omega,s)- C_i \, h_i''(0;\,k,\,\omega,s)\, \rp]
\label{DRAGP}
\eq
where we have assumed $C=a\,b\sim k^n$ with $n \ge 1$ as $k\to 0$ as stated by condition \rf{KT0}.

For a laminar flow, $a$ and/or $b$ must vanish as explained in \S\ref{ORD1}, so that equation \rf{DR} gives $\DR=0$. Again, this result is consistent with the results described in \cite{quadrio-ricco-2011}, where the laminar Poiseuille flow is demonstrated not to be affected by the GSL originated by the travelling waves, since the streamwise parabolic velocity profile and the transverse GSL profile are entirely decoupled. In our model,  the basic equations \rf{UTURB}--\rf{CONT} for a laminar flow are no longer coupled through the turbulent viscosity $\nu_T$, and this suggests reconsidering the basic perturbative assumptions \rf{UPERT}--\rf{WPERT2}, checking their appearance when $a=0$ or $b=0$ or $a=b=0$. It turns out that the right closure of the problem for the laminar case is $b=0$, because in this way the assumptions
\rf{UPERT}...\rf{WPERT2} take the form
\bq
&&u = u_0(y) \\
&&v = 0      \\
&&w = \en\, f(y)\, \ex^{i \theta} 
\eq
($b=0$ in the continuity equation \rf{CONT1} gives $g(y)=0$) and represent a flow where the $x$ and $y$-momentum equations are independent from the $w$ velocity component, whilst this component can be calculated by solving equation \rf{FLAM}, that couples $f$ with $u_0$. 

\section{Results}
\label{RESULTS}

In order to obtain $h''(0)$ and thus the map of the drag reduction $\DR$ as a function of the parameters $k$ and $\omega$, eq. \rf{HEQ2} must be completely characterized by specifying the functions $u_0$ and $\nu_0$, then solved with the relevant boundary conditions on a suitable domain $0\le y\le y_m$, i.e. a range of distances from the wall to a position where the drag reduction effects of the moving wall become negligible. The details of this procedure are presented in the following sections.

\begin{figure}
\begin{center}
\includegraphics[width=0.6\textwidth,angle=0]{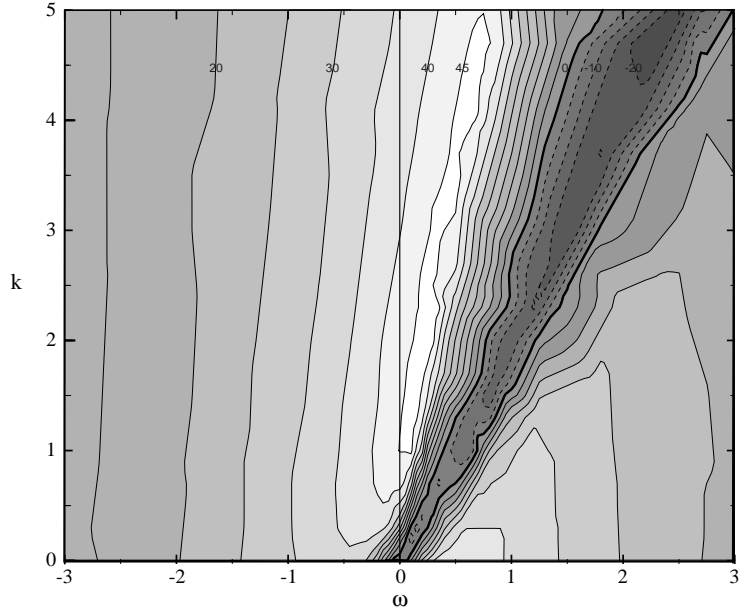}
\caption{Map of drag reduction (percentage) at low $Re$ in the $k$-$\omega$ plane, adapted from \cite{quadrio-ricco-viotti-2009}. The thick contour line marks the neutral level (no drag variation). Friction drag increases within the darker triangular region.}
\label{DNS} 
\end{center}
\end{figure}

The solution procedure is expected to lead to a $\DR(k,\omega)$ map, that requires first to be compared with known results for validation. Quadrio et al. in \cite{quadrio-ricco-viotti-2009} provide an ample dataset created with a comprehensive numerical study. They carried out state-of-the-art Direct Numerical Simulations (in terms of computer code and numerical algorithms, and -- most importantly -- in terms of discretization parameters and solution procedure) to determine the dependence of drag reduction $\DR$ upon $k$ and $\omega$ at $Re=4760$. Fig. \ref{DNS} reports their results, and can be considered as a reference map for comparison. This map clearly exhibits an oblique region near the vertical axis where $\DR$ reaches its maximum value (referred to as {\em hill} in the following) and a triangular region where $\DR$ has a negative minimum, being a region of drag increase ({\em valley}). These regions are separated by an oblique line where $\DR=0$, called the {\em neutral line}.

\subsection{Problem closures and general properties}
\label{clos}

Once $u_0$ and $\nu_0$ are known, the problem at hand is given by eq. \rf{HEQ2} together with its boundary conditions. The most general way to impose these conditions is based on the consideration that the physical meaning is contained in the real part of $h$ and $g$, so the problem may be written in the form
\bq
\cL_2 h&=&0 \label{L1H}\\
\Re\{h(0)\}&=&0 \label{CH} \\
\Im\{h'(0)\}&=&- \frac k b\, \Re\{g(0)\}=0 \label{CG} \\
\Im\{h'(y_m)\}=0 &\hbox{or}& \Re\{h(y_m)\}=0 \label{CM}
\eq
where conditions \rf{CH} and \rf{CG} state that the real parts of $h$ and $g$ vanish at the wall, i.e. $u_1=v_1=0$, whilst condition \rf{CM} states that $g$ or $h$ ($u_1$ or $v_1$) vanishes at the far boundary $y_m$, representing 2 possible closures of the system.
A first study of this system can be made substituting power series for $u_0$ and $\nu_0$ of the kind 
\bq
u_0 = \sum_{n=0}^N a_n\, y^n +O(y^{N+1}) \; ; \;\; \nu_0 = \sum_{n=0}^N b_n\, y^n +O(y^{N+1})
\label{USR}
\eq
and according to these approximations, a similar series for the unknown function $h$:
\bq
h = \sum_{n=0}^N c_n y^n +O(y^{N+1}) . 
\label{HSR}
\eq

The resulting set of equations gives non trivial solutions (where any $c_n$ is non zero) 
%
%
only if $h$ and $g$ at the wall vanish completely (real and imaginary part), so that the problem under study becomes
\bq
\cL_2 h&=&0 \label{L2H}\\
h(0)&=&0 \label{CH0} \\
h'(0)&=&0 \label{CG0} \\
\Im\{h'(y_m)\}=0 &\hbox{or}& \Re\{h(y_m)\}=0 \label{CM0}.
\eq
The two closures expressed by condition \rf{CM0} become in explicit form
\bq
h'(y_m)=s &\hbox{or}& h(y_m)=\im s
\eq
where $s$ is a real number with the meaning of an integration constant.

Several properties of the function $h$ can be revealed by the power series method (see appendix \ref{DRPS}).
The most important are outlined in what follows for the problem closed by condition $h'(y_m)=s$; it is easy to see that these properties remain unchanged even with the second closure, with the exception that the roles of real and imaginary parts $h_r$ and $h_i$ are exchanged, as well as their signs. The power series method gives general results of the kind
\bq
h''(0;k,\omega,s) \propto \,\frac{s}{P(k,\omega; y_m)} 
\label{HS0}
\eq
where $P(k,\omega; y_m)$, having $y_m$ as a parameter, is a $k,\omega$-polynomial whose order depends on the chosen order for the function $h$. The parameter $s$ turns out as a factor owing to the homogeneity of eq. \rf{L2H}, and only affects the amplitude of $h''(0;k,\omega,s)$, that remains unknown anyway because of the perturbative approach.
From \rf{HS0} it is easy to see that $h''(0;k,\omega,s)$ is regular and tends to 0 as $k\to\pm \infty$ and $\omega \to\pm \infty$. This property has just a mathematical meaning, since under this limit the wavelength and period of the wall forcing $L_{wall}$ and $T_{wall}$ tend to 0, which is a violation of the initial hypothesis that the forcing takes place on scales that are large with respect to the turbulent spatial and temporal scales.

After defining the domain $0\le y\le y_m$, it can be seen that the use of a truncated power series for $h$ of order $\ge 5$ is sufficient to reproduce the basic features of the original, DNS-computed map of $\DR$, leading to imaginary parts $h_i''$ that exhibit a oblique hill and a parallel valley separated by a neutral line. Actually, once $h$ is known, a reconstruction of the $\DR$ map may be attempted using eq. \rf{DRAGP}, that becomes here
\[
\DR_r(k,\,\omega) \propto \, \frac 1 k \lp[\, C_r \, h_r''(0;\,k,\,\omega)- C_i \, h_i''(0;\,k,\,\omega)\, \rp] 
\] 
where the constant $s$ is no longer highlighted thanks to its factorization. 
Another general property tells that in the origin of the $k - \omega$ plane $h_i''=0$ whilst $h_r''\ne 0$. This suggests to assume $C_r=0$ since at $(k,\omega)=(0,0)$ the wall motion does not exist and in this case the drag reduction must reduce to zero.
The dependence of $C$ on $k$ and $\omega$ can be expanded in a standard power series of the kind
\[
C = C_0 + C_1 \,k + C_2 \,\omega + C_3 \,k^2 + C_4 \,k \,\omega + C_5 \,\omega^2 + \cdots
\]
that becomes, owing to hypothesis \rf{KT0} about regularity in the origin,
\bq
C = k \,(C_1 + C_3 \,k + C_4 \,\omega + \cdots) .
\label{CEXP}
\eq
The practical use of this expansion requires a truncation, and it can be seen that a linearized form of $C(k,\omega)$, together with a truncated series of order $\ge 5$ for $h$, are sufficient to give an approximated version of the $\DR$ map comparable to the reference one. 
A similar method will be applied in the following sections starting from functions $h$ numerically calculated, and leading to results of better accuracy.

\subsection{Numerical solution}
\label{NUMS}

The search for a numerical expression of the function $h(y)$ starts by considering the original equation \rf{HEQ2} together with the accessory conditions, i.e. the problem \rf{L2H} \ldots \rf{CM0}. The functions $u_0$ and $\nu_0$ which characterize equation \rf{HEQ2} are defined by the method (ii) outlined in \S\ref{ORD0}, starting with a formula for $u_0$ expressed in terms of special functions that accurately approximate the mean velocity profile, which is linear very near to the wall (viscous layer), then gradually changes through the buffer layer to a logarithmic behaviour which is typical of the so-called logarthmic layer.

With reference to the accessory conditions, it must be noted that the numerical approach allows an easier closure for the original problem, which can be imposed at the wall, in the following way:
\bq     
\cL_2 h&=&0 \\
h(0)&=&0  \\
h'(0)&=&0  \\
h''(0)&=&S ,
\eq
where $S$ is the integration constant necessary to the closure. Here $S$ is a constant with respect to $y$, but it may depend on the other parameters, i.e. $S=S(k,\omega)$. 

Then, it is useful to take advantage of homogeneity of eq. $\cL_2 h=0$, which allows to factor out the integration constant $S$. A new normalized function $\tih$ can be introduced:
\bq
h\,(y;\, k,\, \omega,\,S)= S\, \tih(y;\, k,\, \omega) .
\eq
This leads to the new system:
\bq
\cL_2 \tih&=&0 \label{L2HB}\\
\tih(0)&=&0 \label{CH0B} \\
\tih'(0)&=&0 \label{CG0B} \\
\tih''(0)&=&1 \label{CMB} .
\eq
Finally, after determining the numerical solution $\tih$, it is possible to set $S$ in such a way as to 
make the system \rf{L2HB}--\rf{CMB} equivalent to the original one \rf{L2H}--\rf{CM0}. 
To do that, $S$ can be defined in two ways: the first one is
\bq
S=\frac{1}{\tih'(y_m)} \;\hbox{ so that }\; h'(y_m) = S\, \tih'(y_m) = 1
\label{s1st}
\eq
i.e. $\Im\{h'(y_m)\}=0$, which is the first closure of the original system already used for power series solutions.
The second possible definition for $S$ is
\bq
S=\frac{\im}{\tih(y_m)} \;\hbox{ so that }\; h(y_m) = S\, \tih(y_m) = \im
\eq
i.e. $\Re\{h(y_m)\}=0$, which is the second closure of the original system. It is easy to show that definitions of $S$ of the kind $S={m}/{\tih'(y_m)}$ and $S={\im\,m}/{\tih(y_m)}$ where $m$ is a number are equivalent to the proposed ones, since a constant $m$ can always be factorized.

\subsection{Drag reduction}

The numerical solution of the equation for $h(y)$ can be obtained by standard methods (Gear/Adams).  The closure condition of the first or second type, respectively, where $\Im \{h'(y_m)\}$ or $\Re \{h(y_m)\}$ vanish at a given distance $y_m$ from the wall, can now be formally imposed at any $y_m$ in the $y$ domain from 0 to positions far from the wall.

\begin{figure}
\begin{center}
\includegraphics[width=0.8\textwidth,angle=0]{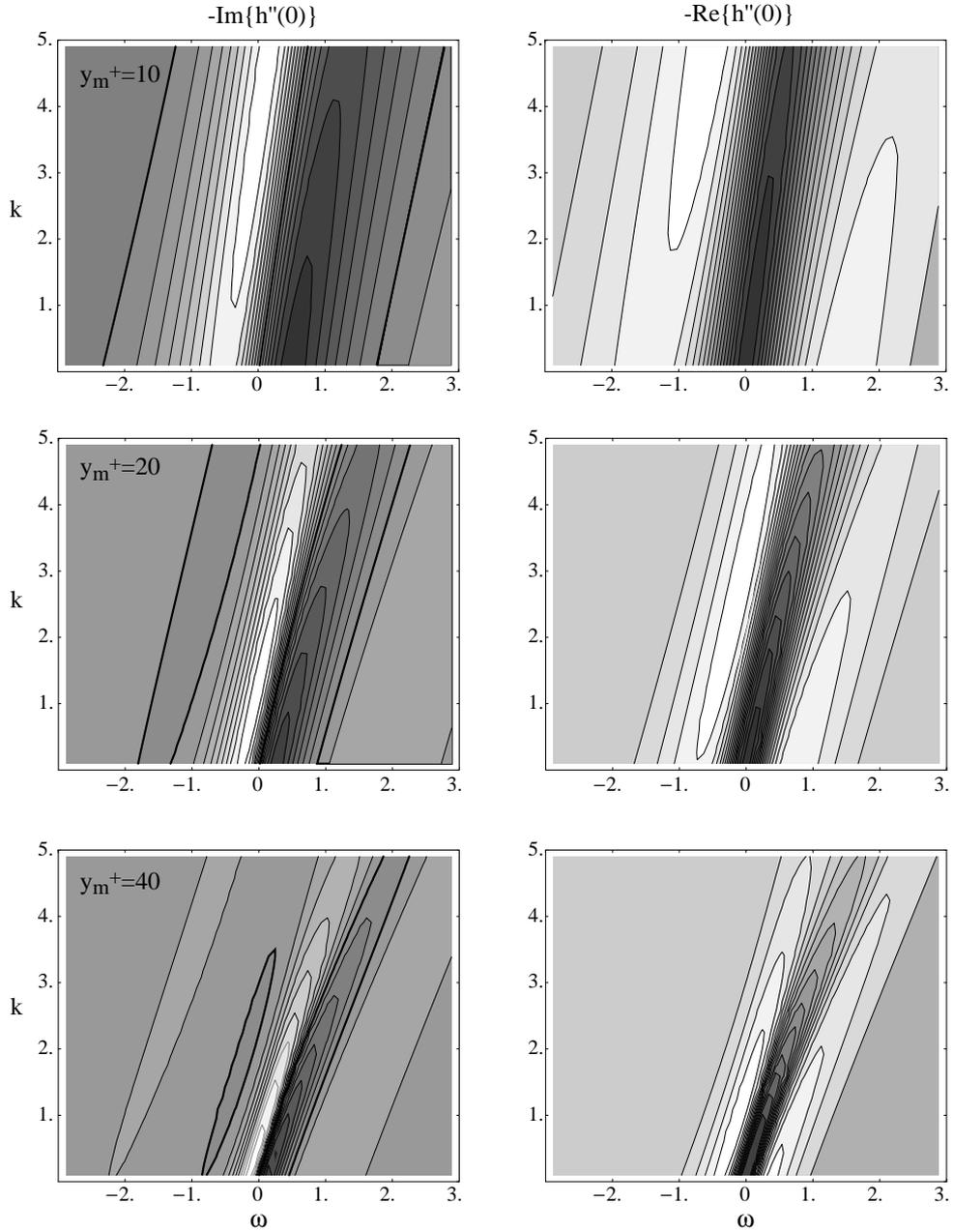}
\caption{Maps of the imaginary (left) and real (right) parts of the function $h''(0;\, k,\, \omega,\,y_m)$, computed for various values of the parameter $y_m$, and with the closure condition of the first kind at $Re=4760$. The neutral line, as all the points where $h''=0$, is evidenced in the maps of $h_i''\,(0;\, k,\, \omega,\,y_m)$ as a thick solid line.}
\label{h-ym} 
\end{center}
\end{figure}

As a first step, the problem at the reference Reynolds number $Re=4760$ is considered with the first closure condition. The function under study is $h''(0;k,\omega,S) = h''(0;\, k,\, \omega,\,y_m)$ because the integration constant $S$ depends on $y_m$ directly, after formula \rf{s1st}. An outline of the results for various values of $y_m$ is shown in fig. \ref{h-ym}. In general, as already revealed by the power series study mentioned in \S\ref{clos}, the numerical solutions too have a real part $h''_r=\Re\{h''(0;k,\omega,y_m)\}$ that does not tend to zero as $(k,\omega)\to(0,0)$ whilst the imaginary part does. Again, drag reduction appears to be better represented by the imaginary part $h''_i=\Im\{h''(0;k,\omega,y_m)\}$, a surface always characterized by a main hill-valley system having minor lateral undulations side by side (it can be verified that it tends to 0 as $k\to\pm \infty$ and $\omega \to\pm \infty$). This can be formally expressed setting $C_r=0$ in the equation \rf{DRAGV}.

\begin{figure}
\begin{minipage}[t]{0.48\textwidth}
\includegraphics[width=\textwidth,angle=0]{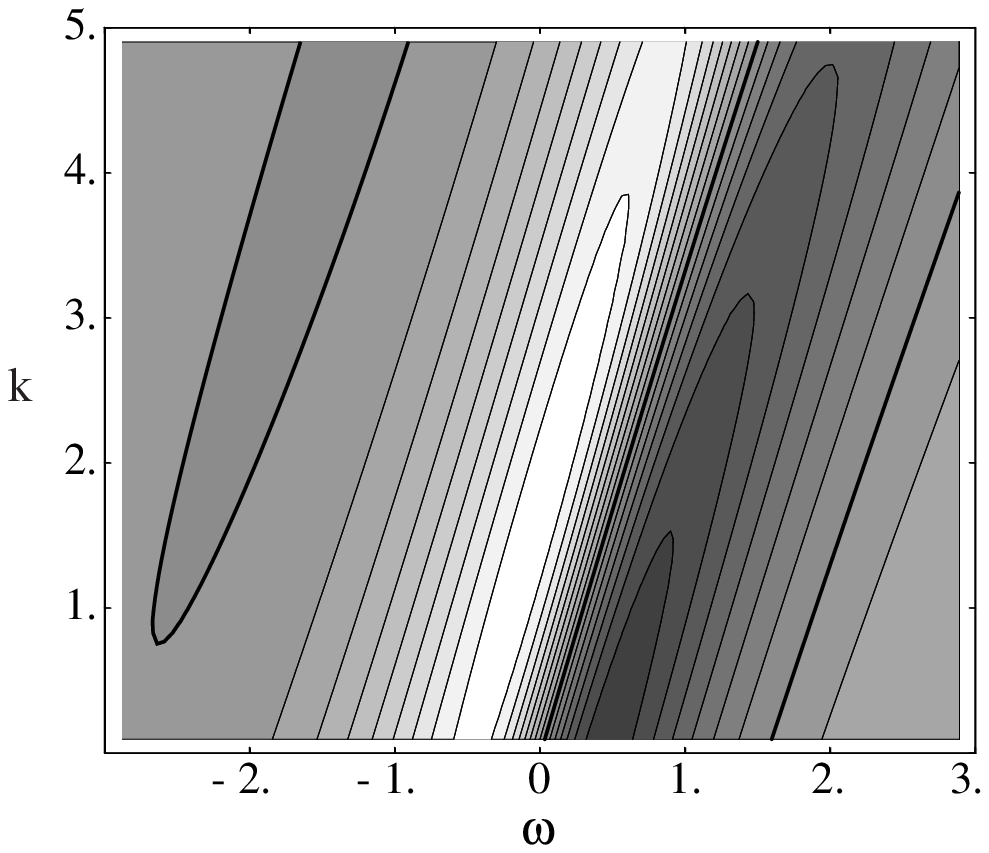}
\caption{Map of $h''_i(y;k,\omega)$, at the optimal $y_m$ and at the reference Reynolds number $Re=4760$.}
\label{hopt} 
\end{minipage}
\hfill
\begin{minipage}[t]{0.48\textwidth}
\includegraphics[width=\textwidth,angle=0]{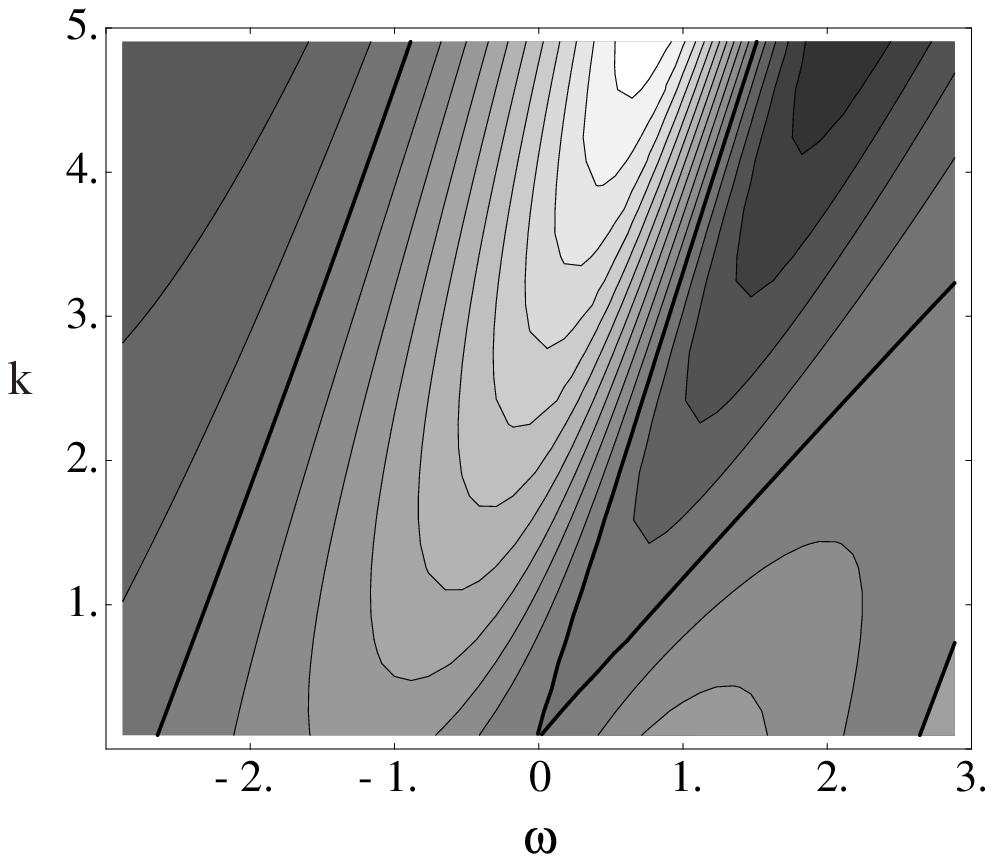}
\caption{$\DR_r$ map approximated by formula $- (C_{3i} k + C_{4i} \omega) \, h''_i(0;k,\,\omega)$ where the coefficients $C_{3i}$ and $C_{4i}$ are  optimized on the $k\omega$ domain under study at the reference Reynolds number.}
\label{drapp} 
\end{minipage}
\end{figure}

At this point, determining the correct value for the parameter $y_m$ still is an open problem, that can be solved by searching for the $y_m$ value such that the neutral line between the hill and the valley in the $k, \omega$ plane is as close as possible to the one in the DNS reference $\DR$ map. This happens for quite a small value of $y_m^+$, namely $y_m^+=27$, confirming that the effects of the forcing are important only in a thin zone adjacent to the moving wall, approximately the size of the buffer region of the mean velocity profile. 
 
The choice of the particular value for $y_m$ value is not critical: we have verified that the following results are almost insensitive to relatively small changes of $y_m$, as long as $y_m$ does not approach the central region of the channel, where they start changing dramatically. Furthermore, it can be seen that a study based on the second closure condition $\Re\{h(y_m)\}=0$ leads to analogous conclusions. The $h''_i$ map with the first closure for $y_m^+=27$ at the reference Reynolds number is shown in fig. \ref{hopt}.

The reconstruction of the $\DR$ map can now be attempted by using equation \rf{DRAGP}, i.e.
\[
\DR_r(k,\,\omega) \propto \, \frac 1 k \lp[\, C_r \, h''_r(0;k,\,\omega)- C_i \, h''_i(0;k,\,\omega)\, \rp]
\] 
(the dependence on $y_m$, which is now a constant, is no longer highlighted). As explained above, the simplest representation of $\DR$ can be obtained by setting $C_r=0$, whilst $C_i(k,\omega)$ must be determined in such a way that
\[
\DR_r(k,\,\omega) = \, - \frac 1 k C_i \, h''_i(0;k,\,\omega)
\]
is as close as possible to the reference map. 
This can be done by expanding $C_i$ in powers of $k$ and $\omega$  after equation \rf{CEXP}, truncating the series at the minimum possible order and imposing the necessary properties in such a way as to  determine the unknown constants, i.e. imposing regularity, $\DR=0$ at $(k,\omega)=(0,0)$ and symmetry $(k,\omega)\leftrightarrow(-k,-\omega)$ since the physics of the traveling wave remains the same in both cases.
The simplest truncation which retains the $k,\omega$-dependency is obtained here by setting $C_n=0$ for $n\ge 5$, what gives $\DR_r \propto -C_i h''_i(0;k,\,\omega)/k = - (C_{1i} + C_{3i} k+ C_{4i} \omega)\, h''_i(0;k,\,\omega)$, that vanishes in the origin. The search for the best coefficient $(C_{1i}+C_{3i} k + C_{4i} \omega)$ fitting the reference DNS map shows that $C_{1i}$ must vanish because of the inherent symmetry of the $\DR$ function for $(k,\omega) \to (-k,-\omega)$ . In this way it is obtained
\[
\DR_r(k,\,\omega) = - (C_{3i}\, k + C_{4i}\, \omega) \, h''_i(0;k,\,\omega) .
\]
By using the function $h''_i$ calculated as in fig. \ref{hopt}, the optimal values of the coefficients on the $k,\omega$ domain under test turn out to be $C_{3i}=0.22, C_{4i}=-0.24$. The resulting map is shown in figure \ref{drapp}, and is in good agreement with the DNS map, the main characteristics of which are qualitatively reproduced: namely, the two-lobed side-by-side pattern with drag reduction and drag increase, and the double local maximum of drag reduction on the $k=0$ axis, corresponding to the temporal oscillating wall emerge clearly from the figure. From a quantitative viewpoint, the neutral line has a slope which is not far from the true one, and the lobes of drag reduction and drag increase have position and extent which are similar to what can be observed in the DNS map. 

\begin{figure}
\begin{minipage}[b]{0.48\textwidth}
\includegraphics[width=\textwidth,angle=0]{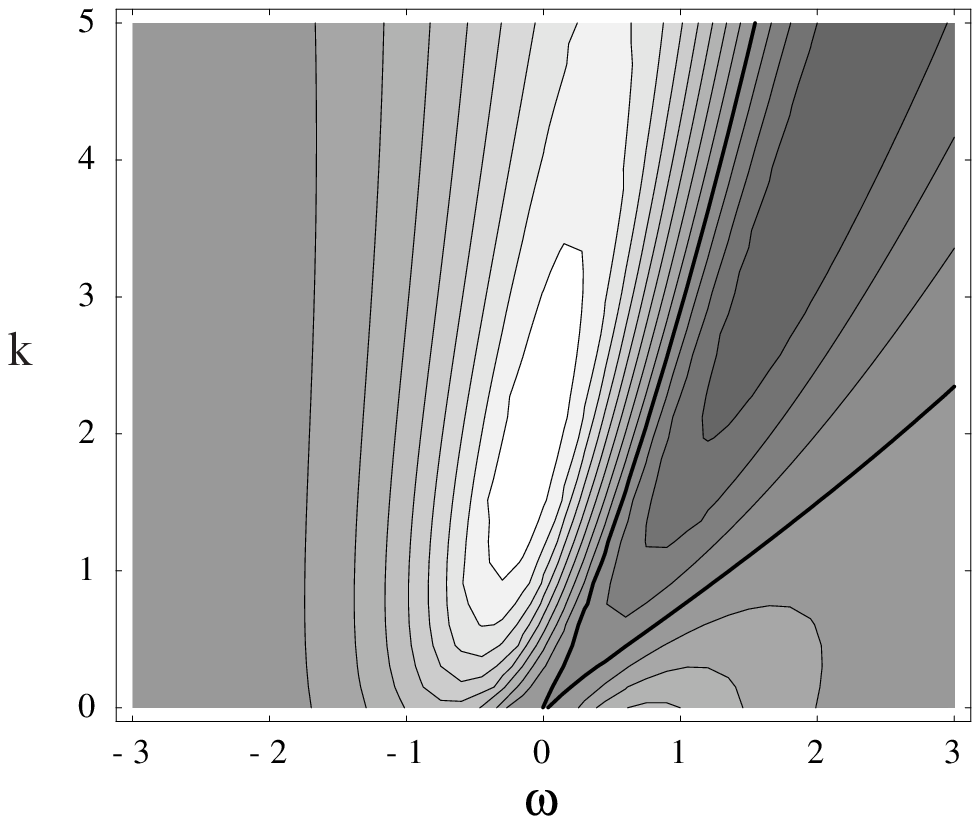}
\caption{$\DR_r$ map approximated by formula 
$\frac 1 k \lp[\, C_r \, h''_r(0;k,\,\omega)- C_i \, h''_i(0;k,\,\omega)\, \rp]$ where the coefficients $C_{i}$ and $C_{r}$ are expressed as 2nd order $k,\omega$-polynomials ratios after formula \rf{PRATIO} and optimized on the $k-\omega$ domain under study at the reference Reynolds number.}
\label{drhigh} 
\end{minipage}
\hfill
\begin{minipage}[b]{0.48\textwidth}
\includegraphics[width=.9\textwidth,angle=0]{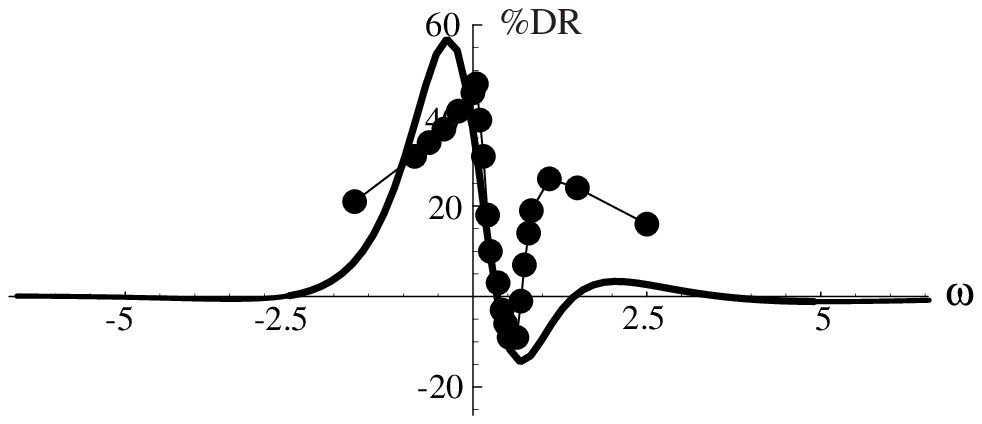}\\[3mm]
\includegraphics[width=.9\textwidth,angle=0]{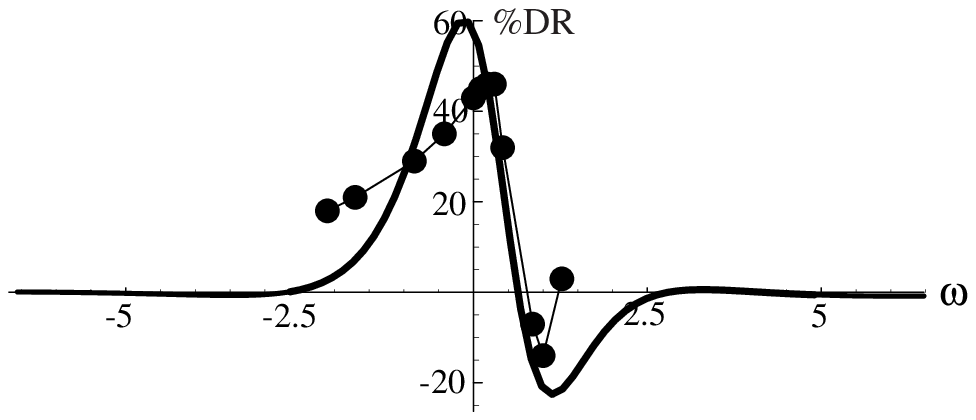}\\[2mm]
\caption{$\DR_r$ values from DNS (dots) and present model (continuous lines) based on the same coefficient $C$ used in figure \ref{drhigh} at $k=1$ (top) and $k=2$ (bottom).}
\label{drks} 
\end{minipage}
\end{figure}

The $C(k,\omega)$ function involved in figure \ref{drapp} is a compact, linearized expression, but nothing prevents us from considering better representations for this factor. Relaxing the assumption $C_r=0$, while keeping  of course the condition $\DR(0,0)=0$ and searching for higher order approximations, a representation of the $\DR$ surface very close to the reference one can be obtained, as reported in fig. \ref{drhigh}. In this case, $C(k,\omega)$ still can be expressed as a truncated power series of the kind \rf{CEXP} as before, but a better representation can be probably written in terms of rational functions:
\bq
C_i &=& k \,\frac{C_{a1} + C_{a2}\, k + C_{a3}\, \omega + C_{a4}\, k^2 + C_{a5}\, k \omega + C_{a6}\, \omega^2 + \cdots} 
{C_{b1} + C_{b2}\, k + C_{b3}\, \omega + C_{b4}\, k^2 + C_{b5}\, k \omega + C_{b6}\, \omega^2 + \cdots} \; , 
\label{PRATIO}
\eq
and $C_r$ is expressed in a similar way. The functions $C_i,C_r$ used in the $\DR$ map of fig. \ref{drhigh} are expressed according to formula \rf{PRATIO} in terms of ratios of simple 2nd-order polynomials in $k$ and $\omega$.

A more quantitative assessment of the results obtained so far can be had by looking at sections of the map taken at fixed values of $k$: for example, fig. \ref{drks} shows two sections of fig. \ref{drhigh}, where it can be seen that there is good agreement in the positions of maxima and minima between DNS results and the prediction of the present model, the only remarkable difference being in the lateral decay rate of the $\DR$ values.

\section{Using the model}
\label{USE}

Our model has been developed, and it has been verified that it is able to produce data which are broadly in line with the available DNS information. As in a typical asymptotic formulation, a part of the information required to solve the problem has been introduced as external data from the underlying physics, namely the domain width $y_m$ and the related third integration constant $s$ or $S$. It is worth to recall here that $y_m$ is the range of distances from the wall to a position where the drag reduction effects of the moving wall become negligible, whereas $s$ or $S$ have the meaning of an accessory condition in addition to the standard no-slip condition at the wall. 
Another effect of the asymptotic approach is the appearance of the multiplicative function $C(k,\omega)/k$ in the equation \rf{DRAGV} for the drag reduction, a function that cannot be intrinsically determined at the first level of approximation on the small nondimensional forcing amplitude $\en$. Even if the basic properties of the $\DR$ map can be represented by the $h''(k,\omega)$ surface, the function $C(k,\omega)$ was shown to be useful for a "calibration" of the model on the available information, approximating it as a first choice in linearized form. 

It is thus of interest to use the model to predict the behavior of drag reduction beyond the available DNS information, which is limited to low values of both the wavenumber $k$ of the traveling waves, and of the Reynolds number $Re$ of the longitudinal flow. In the following we will use our model to investigate the trend of variation of turbulent drag reduction when $k$ is increased, and -- more interestingly -- when $Re$ is increased. This is the ultimate aim of the present paper.

\subsection{Drag reduction at higher values of $k$ and $\omega$}
\label{HIGHK}

A first attempt at using the predictive capabilities of the present model is devoted to investigating the behaviour of turbulent drag reduction at high values of $k$. In the literature, at the reference value $Re=4760$ there are no such data available for $k>5$ and $|\omega|>3$. This first predictive step is intermediate and serves us the purpose of testing the model on a region of the parameter space for which it has not been calibrated, but for which DNS data can still be produced to evaluate it. We have then purposely run 18 additional DNS of turbulent plane channel flow at $Re=4760$, similar to what has been done in \cite{quadrio-ricco-viotti-2009}. The DNS code, described in \cite{luchini-quadrio-2006}, uses Fourier discretization in the homogeneous wall-parallel directions and compact, IV-order explicit finite difference schemes in the wall-normal direction, and employs a partially implicit time integration method. The computational domain is identical to that employed in the original study: $L_x=6 \pi h$, $L_y=2h$ and $L_z=3 \pi h$. The spatial resolution is $N_x=320$ and $N_z=320$ Fourier modes, and $N_y=161$ points. The main addition to the available data is the range of wavenumbers and frequencies that are investigated, which spans the interval $10<k<30$ and $3 < |\omega| < 15$. 

\begin{figure}
\includegraphics[width=0.32\textwidth,angle=0]{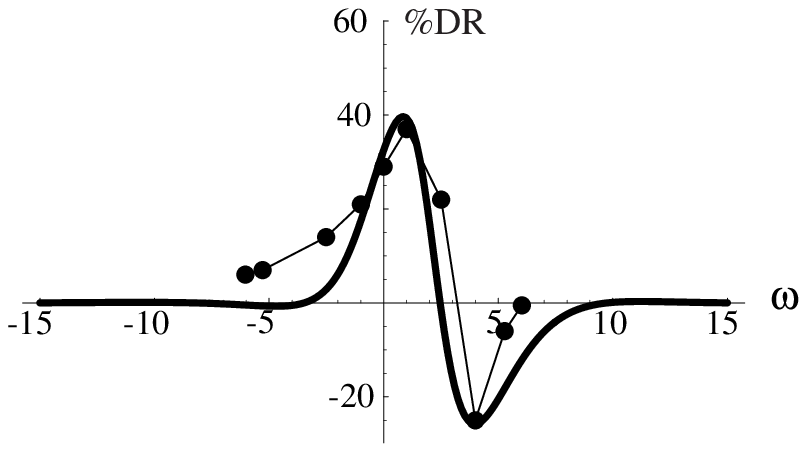}
\hfill
\includegraphics[width=0.32\textwidth,angle=0]{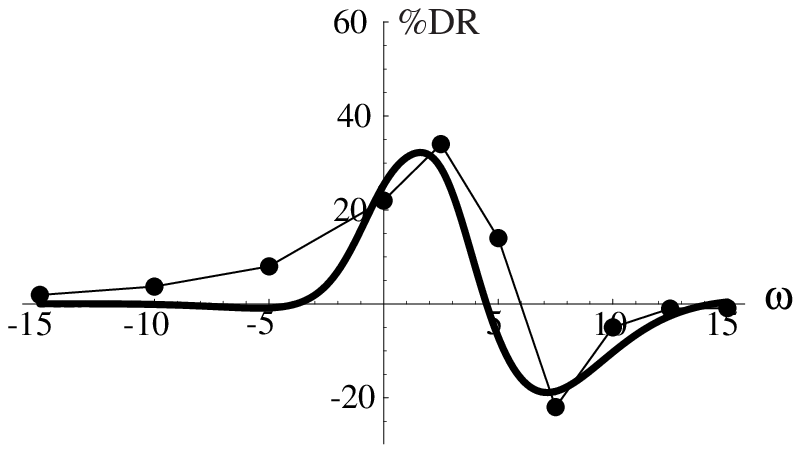}
\hfill
\includegraphics[width=0.32\textwidth,angle=0]{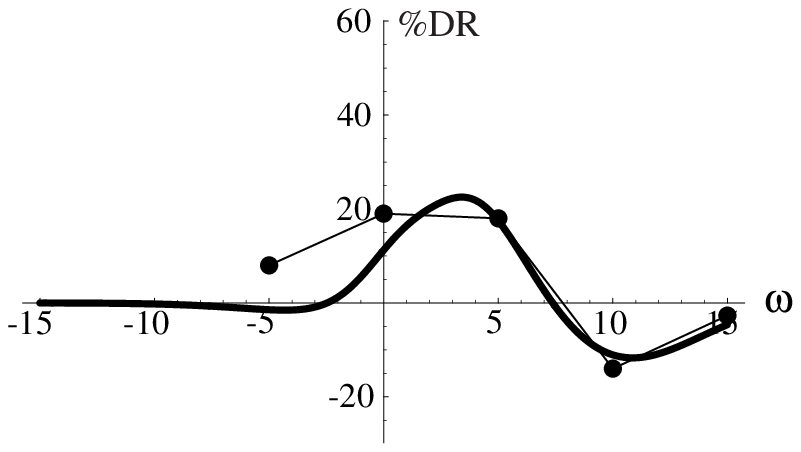}
\caption{$\DR_r$ values from DNS (dots) and present model (continuous lines) based on the same coefficient $C$ used in figure \ref{drhigh} at $k=10$, 20 and 30 from left to right.}
\label{drkl} 
\end{figure}

Owing to the relatively limited number of new data points available, it is difficult to build a reliable two-dimensional map. Figure \ref{drkl} thus compares DNS data (dots) and predictions from the present model (lines) at constant $k$ as a function of $\omega$. Again, the positions of maxima and minima are in good agreement between DNS data and present model. Overall, both results indicate that the hill-valley structure remains elongated and has a rather slow decay in the $k-\omega$ plane.

\subsection{Drag reduction at higher values of the Reynolds number}
\label{HIGHRE}

Applying the present model to investigate the turbulent drag reduction for high values of the Reynolds numbers is obviously the main thrust of the present study. Unfortunately, this is prevented by the problem of determining the coefficient $C$ at these Reynolds numbers. However, another interesting test can be carried out, i.e. computing the function $h''(0;k,\omega)$ for larger and larger $Re$. This can be done rather easily by for example keeping the first closure condition at the optimal value $y_m^+=27$. In this section only the maps for the imaginary part of $h''$ will be considered, but it can be shown that $h''_r$ has a similar scaling.

\begin{figure}
\begin{center}
\includegraphics[width=0.9\textwidth,angle=0]{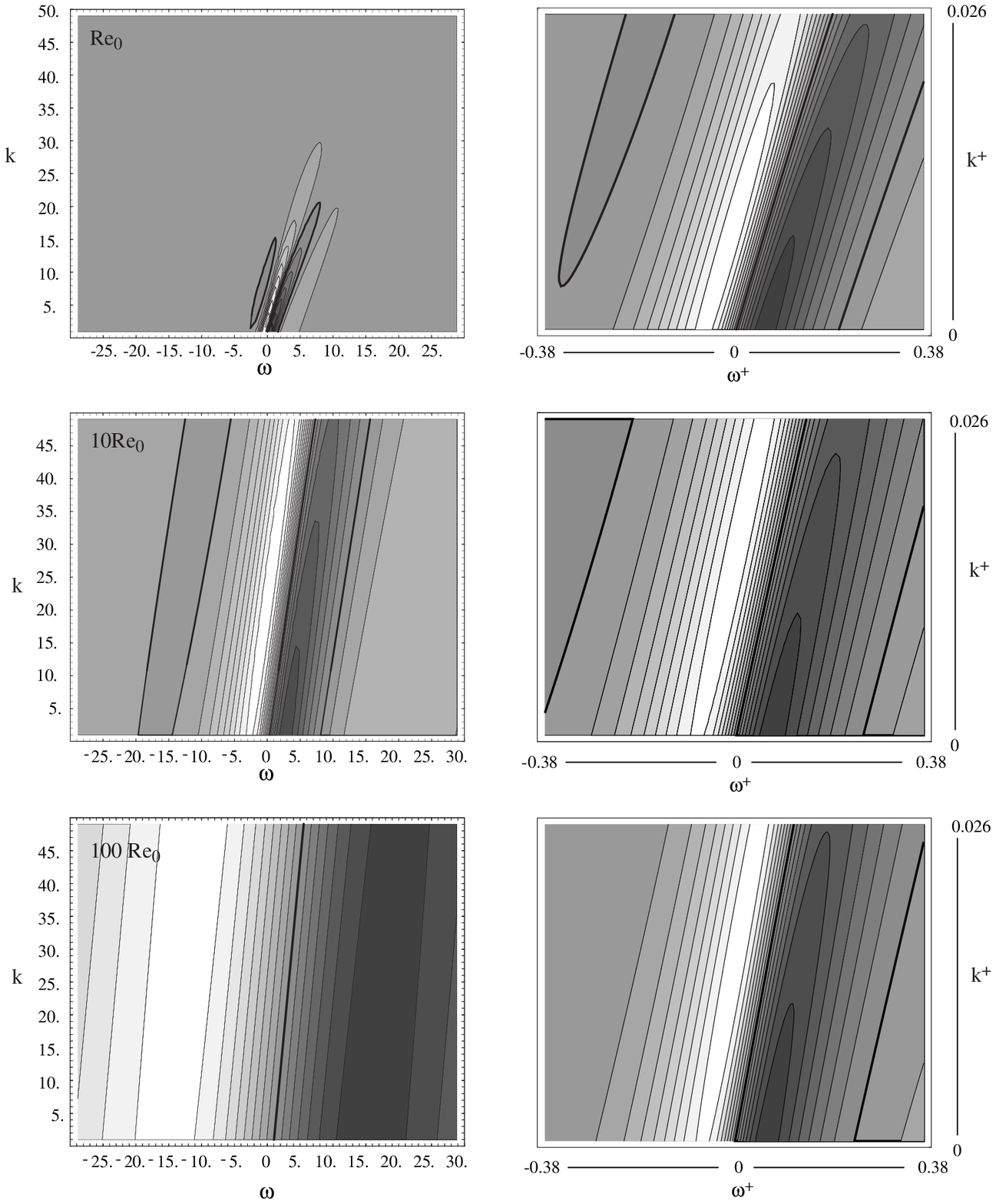}
\caption{Maps of $h''_i(0;k,\omega)$ with $y_m^+=27$ at Reynolds numbers $Re_0=4760$, $10Re_0$ and $100 Re_0$. Left: outer units, plots over the domain $-30<\omega<30$, $0<k<50$. Right: inner units, plots over the domain $-0.38<\omega^+<0.38$, $0<k^+<0.026$ (see the text).}
\label{hi-r} 
\end{center}
\end{figure}

Figure \ref{hi-r} shows the $h_i''$ maps computed with our model at 3 values of the Reynolds number, i.e. the reference one $Re_0=4760$ (top), and the values $10 Re_0$ (middle) and $100 Re_0$ (bottom). Focusing first on the left column, the top plot is an extension of fig. \ref{hopt} to the wider domain $-30<\omega<30$, $0<k<50$ and has been already discussed. As $Re$ increases, the general shape of the map over the same domain seems to change only weakly, if exception is made for an evident increase of its size. This means that the characteristic time and space scales shrink as the Reynolds number increases. 

Since we are concerned with near-wall turbulence, it is reasonable to suppose for the phenomenon to obey an inner, viscous scaling. This is verified in the right column of figure \ref{hi-r}, where the same maps are translated in wall units and are shown to keep their general size across a 100-times increase in Reynolds number. In particular, the right column is obtained by choosing a domain $-0.38<\omega^+<0.38$, $0<k^+<0.026$ in wall units, obtained from the conversion at $Re=Re_0$ of the original one used in figures \ref{DNS} to \ref{drhigh}, that was $-3<\omega<3$, $0<k<5$ in outer units. Then, all the maps of the left column are plotted again over this new domain in wall units, for all the values of $Re$, evidencing in this way the morphologic analogies. Clearly, the inner units maps are not identical, so that a strict Reynolds-invariance property can be excluded; but a property of slow Reynolds-scaling could be revealed by the present analysis.

\begin{figure}
\begin{minipage}{0.25\textwidth}
\begin{tabular}{@{}rl@{}} 
$Re/Re_0$  & $F(Re)/F(Re_0)$ \\
\hline
1 	& 1 \\
2 	& 0.974 \\
4	& 0.949 \\
10 	& 0.919 \\
20	& 0.899 \\
40	& 0.880 \\
100	& 0.858 \\
200	& 0.844 \\
\hline
\end{tabular}
\caption{Normalized amplitude factors vs. Reynolds number.}
\label{tab}
\end{minipage}
\hfil
\begin{minipage}{0.7\textwidth}
\includegraphics[width=\linewidth]{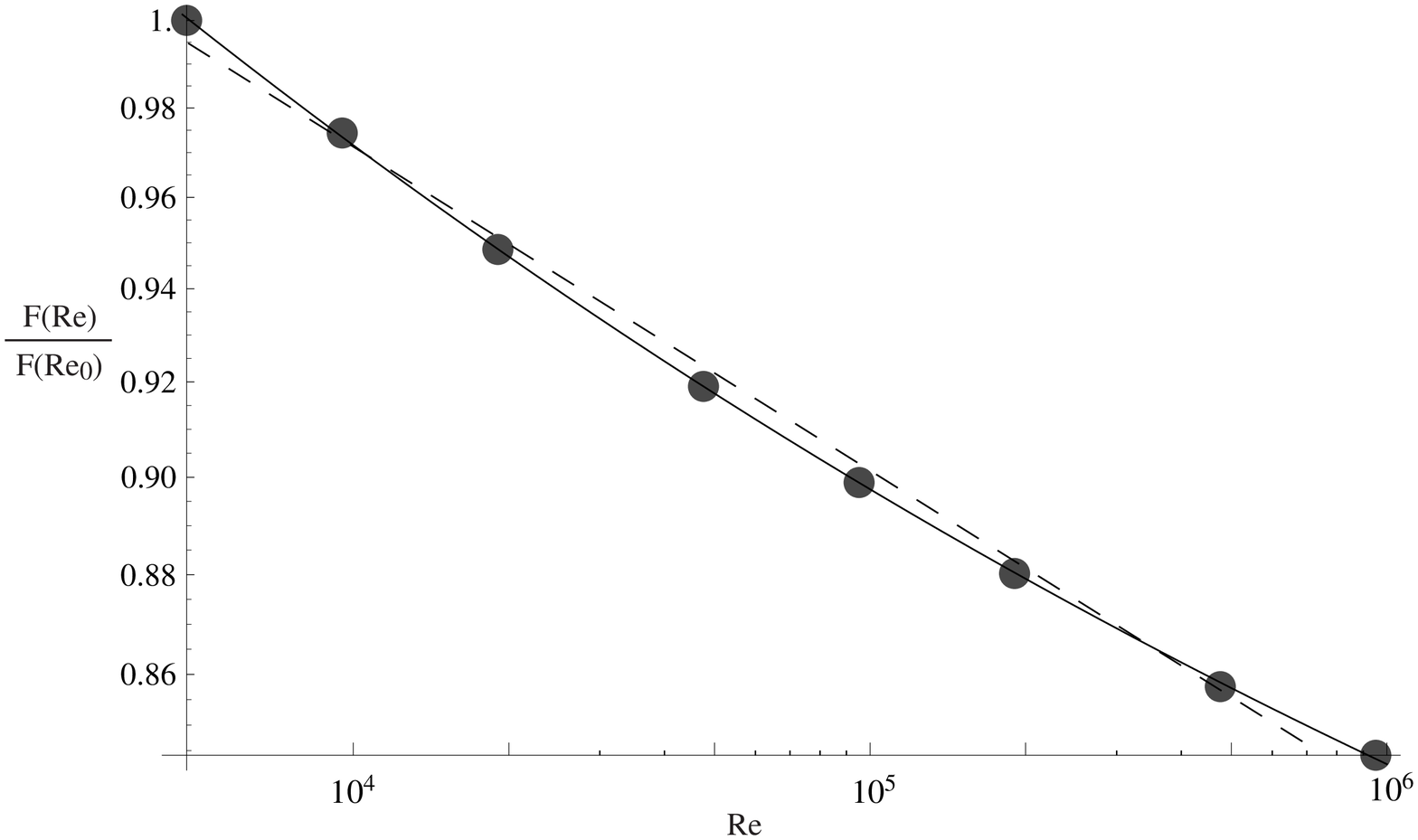}
\caption{Computed decrease of maximum drag reduction as reported in table \ref{tab} (dots).  The lines are two fits for these data: dashed line is after eq. \rf{fit1}, and solid line is after eq. \rf{fit2}.}
\label{drmax}
\end{minipage}
\end{figure}

Of course this perturbative analysis cannot give any indication on the absolute levels of drag reduction that can be achieved. Of particular interest, however, is to investigate how the best performance at low $Re$ degrades at higher $Re$, and how the values of $k$ and $\omega$ defining the most performing traveling wave are observed to change while $Re$ increases.
In order to compare results obtained at different Reynolds numbers, a proper amplitude factor must be defined for the $h''_i$ maps: since $\DR(k,\,\omega) \propto \, C \, h''(0;\,k,\,\omega)/k$ after equation (\ref{DRAGV}), here $F(Re) = h_{i,max}''/k_{i,max}$ is selected as a factor for the present purpose ($k_{i,max}$ is the maximum locus), i.e. a quantity that should give information about the trend of $\DR$ as function of $Re$. The results are shown in table \ref{tab} and figure \ref{drmax}, where the values of $F(Re)$ normalized on the reference value obtained at $Re_0=4760$ are presented: actually, the trend exhibits a slow Reynolds-scaling, and raises a new question, about the existence of an asymptotic limit for large Reynolds numbers. 

Figure \ref{drmax} shows also two possible fits for the trend of the calculated points. The first fit, plotted with a dashed line in the figure, as function of $Re_\tau$ is:
\bq
G(Re_\tau) = F(Re)/F(Re_0) = \alpha\, Re_{\tau}^\beta
\label{fit1}
\eq
where $\alpha=1.31, \beta=-0.0433$. The second fit, plotted with a continuous line in the figure, as function of $Re_\tau$ is:
\bq
G(Re_\tau) = F(Re)/F(Re_0) = \gamma + \alpha\, Re_{\tau}^\beta
\label{fit2}
\eq
where $\gamma=0.682, \alpha=0.945, \beta=-0.171$. Judging from the present data, the second fit appears to be better than the first, but it requires the determination of 3 parameters instead of 2, and implies the existence of a finite, non-vanishing asymptotic limit.

The present result confirms, first of all, the expected decrease of the performance of the travelling waves with increasing Reynolds number. This is of course a very reasonable effect from the physical standpoint, since it reflects the gradual decrease of importance of the viscous near-wall cycle in the overall turbulent flow, which is more and more influenced by the large-scale events originating in the outer (logarithmic) region \cite{laadhari-skandaji-morel-1994}. Although no information is available for the travelling waves at $Re_\tau>400$, we can use the information available for the spanwise-oscillating wall, which is a simpler technique driven by a similar physics and has been investigated up to $Re_\tau=1000$. Touber \& Leschziner \cite{touber-leschziner-2012} for example report LES and DNS information to substantiate the claim that maximum drag reduction decreases with a power law $\sim Re_\tau^{-0.2}$. A similar result was proposed earlier by \cite{choi-xu-sung-2002}, although on the basis of a dataset computed for $100 < Re_\tau < 400$. These results are consistent with our findings, although our rate of decay, computed on the entire dataset available, seems to be slower, with an exponent of -0.043 only. This rate, computed on a smaller dataset for $Re < 10 Re_0$, turns out to be only slightly faster, with an exponent of -0.051. We notice, moreover, that our data lend clearly better support to a decay law of the type \rf{fit2}. This kind of fit would be also more in line with the claim by Iwamoto \cite{iwamoto-etal-2005} that the decay of the maximum drag reduction is a low-Reynolds effect, or at least that at high $Re$ the performance decay becomes very slow.

By its very nature the present approach, being based on low-$Re$ information, is unable to account for a possible major change with $Re$ of the layout of the drag reduction map in the parameter space. That said, the locus $(\omega,k)_{i,max}$ of the maximum values of $h_{i,max}''$ considered above seems to be mildly dependent upon the Reynolds number. Actually, in the range $Re_0 \le Re \le 200 Re_0$ the pulsation varies in the range $0 \le \omega^+_{i,max} \le 4 \cdot 10^{-5}$, which is in the order of the accuracy of the ODE solver employed, whereas the wavenumber shows a very small decrease, namely $0.0089 \ge k^+_{i,max} \ge 0.0081$.

One last comment is in order with respect to the net energy savings brought about by the travelling waves. Drag reduction contributes to energy savings through reduction of the pumping power, but the energetic cost of control must be considered too. There is information available \cite{ricco-quadrio-2008} only for the oscillating-wall case, where the power required to oscillate the wall in a plane channel flow is reported to slightly decrease with $Re_\tau$, being proportional to $Re_\tau^{-0.136}$. This is simply the effect of assuming $Re = 15.43 Re_\tau^{1.136}$ as in \cite{pope-2000}, since in outer units the required power is constant. It follows that the small decrease of drag reduction implies a corresponding small decrease in the net savings. This conclusion is expected to hold for the travelling waves too, on the basis of the validity of the laminar GSL solution to describe the forcing-induced laminar oscillating transverse velocity profile.

\section{Conclusions}
\label{CONCLUSIONS}

The present paper has introduced an innovative perturbation method aimed at predicting the turbulent drag reduction characteristics of the streamwise-travelling waves of spanwise wall velocity for the geometry of a plane channel flow. 

The method, based on a perturbative approach, considers the Reynolds-averaged Navier--Stokes equations in the simple geometry of an indefinite plane channel flow. In addition to the usual streamwise mean flow, a spanwise oscillating flow with zero mean value induced by the wall forcing is present. A perturbation expansion is applied to the mean streamwise and spanwise flow, as well as to the turbulent viscosity profile. Under the fundamental assumption that the spanwise flow is small with respect to the streamwise flow, the perturbative problem expressed at zero-th order corresponds to the standard turbulent channel flow, whereas at first-order one obtains a new equation set, which contains the spanwise flow and its interaction with the streamwise flow. Our interest is focused on the wall derivative of the streamwise velocity profile, a quantity that is easily related to the turbulent friction. 
The modification of this derivative as an effect of the wall forcing is then computed in the asymptotic frame.

The procedure is first used to reproduce the available information on the turbulent drag reduction brought about by the traveling waves. This information describes how drag reduction depends on the wavenumber and temporal frequency of the waves, and was previously computed by Direct Numerical Simulation of the full Navier--Stokes equations. In the present model, a function appearing as a factor in the $\DR$ formula was expressed in two possible ways, at first as a linearized expression, then as a more accurate formula, in both cases this function was used to improve the quality of results, and the relevant coefficients have been obtained by fitting the reference DNS map on a reference $k,\omega$ domain. 

Once the procedure has been wholly established, including its parameters, it is used to produce new information. It is first tested on a new dataset, purposely produced by DNS during the present work, designed to explore the drag reduction induced by the waves at low values of $Re$ but high values of $k$ and $\omega$. By keeping the fit coefficients computed on the smaller $k,\omega$ domain of reference, our method is shown to yield a good prediction. 

Lastly, the main question that motivates the present method is addressed, and the variation of the maximum drag reduction brought about by the travelling waves when the value of $Re$ increases is predicted. This prediction instead has been obtained avoiding the use of fitted coefficients. 
Compared to the baseline value $Re_0$, the available literature information is limited to $2 Re_0$ for the streamwise-travelling waves, and up to $5 Re_0$ for the spanwise-oscillating wall. Here we can easily reach more than $100 Re_0$, so that the two-decades span allows to draw some clear conclusions about a true high-$Re$ trend. Our findings are, first of all, that maximum drag reduction decreases, as expected on the basis of physical considerations, since the near-wall layer becomes less and less influential on the whole turbulence dynamics as $Re$ grows. The important result, however, is that our data suggest a significantly milder decay when compared to the available predictions of the kind $\sim Re_\tau^{-const.}$, which are however limited by the very small extent by which $Re$ has been increased. Instead of the commonly reported decay $\sim Re_\tau^{-0.20}$, we suggest a much slower decay $\sim Re_\tau^{-0.04}$. When an asymptotic value for the drag reduction at large $Re$ is allowed by the chosen fit, our data show a better agreement, thus implying that drag reduction does not decrease below a certain threshold as the value of $Re$ increases.

The predictive procedure described in this paper undoubtedly contains a number of significant and critical assumptions. Just to name the most critical one, we remind the reader of the Boussinesq hypothesis; of the assumption that the amplitude of the transversal velocity waves is small compared to the longitudinal velocity scale; of the assumption that (space and time) scale separation exists between the waves and the turbulent flow. That said, however, the scenario that emerges from using the procedure is generally consistent, as far as predictions can be critically evaluated against available data. 
When the procedure is used to make true predictions in a regime where no information is available, the conveyed message is reasonable and points to a high-$Re$ scenario that is less pessimistic compared to what is generally thought of on the basis of the available information.
 Without counting too much on the exact validity of the predicted scenario, we would like to offer this more optimistic high-$Re$ view to the flow control community, as a further motivation to intensify the efforts towards understanding what really happens to wall-based turbulent drag reduction techniques when the values of the Reynolds number become significantly high and reach application-level. 

\bibliographystyle{zamm-title}
\bibliography{mq}


\appendix
\vspace*{12mm}

{\bf \large Appendix}

\section{Drag reduction from power series method}
\label{DRPS}

The simplest attempt to solve problem \rf{L2H}...\rf{CM0} without loss of physical meaning makes use of series of the kind \rf{USR} 
for $u_0$, $\nu_0$, truncated at the 3rd order in $y$. The relevant polynomial coefficients $a_n$ and $b_n$ can be fit on the numerous literature data.  
This level of approximation permits to write $h''(0)$ in a closed form of acceptable size, 
but it is limited to low values of $y$ in the order of $0<y^+ \alt 50$, since it may become inaccurate over a large domain.
As regards the unknonwn function $h$, it is easy to see that the solutions of the problems based 
on the power series \rf{HSR} for $h$ truncated at the $N$-th order leads to trivial results until $N\le 4$ with either closure condition, 
$\Im\{h'(y_m)\}=0$ or $\Re\{h(y_m)\}=0$. 


The first interesting case is obtained at order 5 in $y$: here the problem closed by condition
$\Im\{h'(y_m)\}=0$, formally expressed as $h'(y_m)=s$ where $s$ is a real number, gives
\bq
h''(0;k,\omega,s)=  - \frac {24 \,s \, {\nu }^3 } { y_m} \,\frac{1}{\alpha + \beta \, \omega  + \gamma_1 \,k + \gamma_2 \, k^2 } 
\label{HS05}
\eq
with
\bq
\alpha &=& 6\,b_1^3\,y_m^3 - \left(8\,b_1^2\,y_m^2 + 12\,b_1\,b_2\,y_m^3 \right) \,\nu 
+ \left( 12\,b_1\,y_m + 8\,b_2\,y_m^2 +  6\,b_3\,y_m^3 \right) \,{\nu}^2 - 24\,{\nu }^3 \\
\beta &=& 4\,\im \,y_m^2\,{\nu }^2 -4\,\im \,b_1\,y_m^3\,\nu \\
\gamma_1 &=& - \im \,a_2\,y_m^3\,{\nu}^2\\
\gamma_2 &=& 3\,b_1\,y_m^3\,{\nu }^2 - 4\,y_m^2\,{\nu }^3 .
\eq
Here $h''(0;k,\omega,s)$ tends to 0 as $k\to\pm \infty$ and $\omega \to\pm \infty$, a property outlined in \S\ref{clos} that holds also for higher order approximations. 
In equation \rf{HS05}, the parameter $s$ appears as a factor thanks to the homogeneity of the $h$-equation \rf{L2H}, and affects only the general amplitude of $h''(0;k,\omega,s)$, which remains unknown because of the perturbative approach.
Thus, the real and imaginary parts of $h''(0;k,\omega,s)/s$ can be easily plotted after setting a value for $y_m$,  
and even in this simple model some features of the $\DR$ map appear to be already present: actually, the imaginary part exhibits a oblique hill and a parallel valley of symmetric heigth separated by a neutral line where $h_i''(0)=0$. The equation of this line
can be deduced from eq. \rf{HS05}, and in this simple approximation it turns out to be a linear relation,
\bq
k = \frac {4 (\nu-y_m b_1)} {a_1 y_m \nu}\, \omega \, ,
\eq
where the steepness depends on $y_m$ progressively more slowly as $y_m$ grows, revealing the dependence on the outer boundary location $y_m$.
The hill-valley structure is bounded for very large values of $k$ and $\omega$, i.e. 
this structure vanishes in the regions of the $k \omega$ plane far from the origin, after eq. \rf{HS05}. 
At the same order in $y$, i.e. with an approximation for $h$ up to $y^5$, but with closure by condition
$\Re\{h(y_m)\}=0$, formally expressed as $h(y_m)=\im\,s$ where $s$ is a real number, similar results are found, 
except that the roles of real and imaginary part are exchanged, as well as their signs.


The approximation for $h$ truncated at the next level, i.e. 6th order in $y$, can be investigated to test the dependence
of the results on the approximation order. 
In this case the problem closed by condition $\Im\{h'(y_m)\}=0$ gives an expression for $h''(0;k,\omega,s)$ of the kind
\bq
h''(0;k,\omega,s) \propto 
\frac {s \,\nu^4}{y_m}\,
\frac{1}{\alpha +\beta_1\ \omega + \beta_2\, \omega^2 + \gamma_1\, k +\gamma_2\, k^2 + \gamma_4\, k^4} .
\eq
Here the coefficients $\alpha, \beta_1, ... \gamma_4$ are quite cumbersome, but many properties of the previous order
remain the same, as $h''(0;k,\omega,s)\to 0$ as $k\to\pm \infty$ and $\omega \to\pm \infty$. 
As above, $s$ can be factorized without loss of generality, and the real and imaginary parts of $h''(0;k,\omega,s)/s$ can be easily plotted.
Even in this case, the imaginary part $h''_i(0)$ exhibits a oblique hill and a parallel valley, but unlike the previous order this hill-valley structure is non-symmetric, with an hill height greater than the valley depth. This structure vanishes fast outside a $k\,\omega$ domain 
which is remarkably smaller than in the previous case. Furthermore, the transition between 
the hill and the valley is smoother.
It can be shown that the 6th order problem with the second closure $\Re\{h(y_m)\}=0$ still gives similar results.


A study of the successive approximations at higher orders becomes rapidly complicated, and it is probably a slowly converging process, requiring at each order a careful check of the dependence of function $h$ on the approximations used for $u_0$ and $\nu_0$ and on the extension of the $y$-domain. 
This suggests to consider a different approach to the original problem, leading to the numerical method exposed in \S\ref{NUMS}.
However, the power series study suggests also that some properties have a general validity; for example, this is the case of the limit $h''(0) \to 0$ for large $|k|$ and $|\omega|$ and of the values $h''_i(0)=0$ and $h''_r(0)\ne 0$ in the origin of the $k \omega$ plane. The general shape of $h''(0)$ at the $N$-th order of approximation turns out to be 
\bq
h''(0;k,\omega,s) \propto \,\frac {s \,\nu^N}{y_m}\,\frac{1}{P_N(k,\omega; y_m)} 
\eq
where $P_N(k,\omega; y_m)$, having $y_m$ as a parameter, is a $k,\omega$-polynomial whose order depends on the index $N$.

The power series approach may serve also as a guide for introducing a reconstruction method of the $\DR$ map, starting from 
equation \rf{DRAGP} rewritten as
\[
\DR_r(k,\,\omega) \propto \, \frac 1 k \lp[\, C_r \, h''_r(0;k,\,\omega)- C_i \, h''_i(0;k,\,\omega)\, \rp] 
\] 
after omitting the dependence on the parameter $s$.
As said above, in the origin of the $k \omega$ plane the real and imaginary parts of function $h''(0)$ have different behaviours, 
namely $h''_i(0)=0$ and $h''_r(0)\ne 0$. Thus, a simple, proper assumption can be $C_r=0$, in such a way as to set $\DR_r=0$
at $(k,\omega)=(0,0)$, when the wall doesn't oscillate.
Then, the dependence of $C$ on $k$ and $\omega$ can be expanded in a standard power series of the kind
\[
C = C_0 + C_1 \,k + C_2 \,\omega + C_3 \,k^2 + C_4 \,k \,\omega + C_5 \,\omega^2 + \cdots
\]
and accounting for the condition \rf{KT0} about the regularity of $\DR$ as $k\to 0$, $C_0$ and $C_2$ must vanish, so the series can be rewritten as
\bq
C = k \,(C_1 + C_3 \,k + C_4 \,\omega + \cdots) .
\eq
In the practical use, this expansion can be truncated, and the coefficients $C_i$ can be fit on known data,
satisfying the symmetry $(k,\omega)\leftrightarrow(-k,-\omega)$, in order to obtain a drag reduction map of the kind
\[
\DR_r(k,\,\omega) \propto \, (C_{1i} + C_{3i} \,k + C_{4i} \,\omega + \cdots) \, h''_i(0;k,\,\omega)\, .
\]

\end{document}